\documentclass[amsmath, amssymb, preprintnumbers, showpacs, showkeys, aps, prl, twocolumn]{revtex4-1}

\usepackage{graphicx}
\usepackage{float}
\usepackage{color}

\begin{document}

\title{A Non-Centrosymmetric Superconductor with a Bulk 3D Dirac Cone Gapped by Strong Spin Orbit Coupling}

\author{Mazhar N. Ali$^{1}$}
\email{mnali@princeton.edu, rcava@princeton.edu}
\author{Quinn Gibson$^1$}
\author{Tomasz Klimczuk$^{2,3}$}
\author{R. J. Cava$^1$}

\affiliation{$^1$Department of Chemistry, Princeton University, Princeton New Jersey 08544, USA.}
\affiliation {$^2$Faculty of Applied Physics and Mathematics, Gdansk University of Technology,
Narutowicza 11/12, 80-233 Gdansk, Poland}
\affiliation {$^3$Institute of Physics, Pomeranian University, Arciszewskiego, 76-200
Slupsk, Poland}

\date{\today}

\begin{abstract}

Layered, non-centrosymmetric, heavy element PbTaSe$_2$ is found to be superconducting. We report its electronic properties accompanied by electronic structure calculations. Specific heat, electrical resistivity and magnetic susceptibility measurements indicate that PbTaSe$_2$ is a moderately coupled, type-II BCS superconductor (T$_c$ = 3.72 K, Ginzburg-Landau parameter $\kappa$ = 14) with an electron-phonon coupling constant of $\lambda_{ep}$ = 0.74. Electronic structure calculations reveal a single bulk 3D Dirac cone at the K point of the Brillouin Zone derived exclusively from its hexagonal Pb layer; it is similar to the feature found in graphene except there is a 0.8 eV gap opened by spin-orbit coupling. The combination of large spin-orbit coupling and lack of inversion symmetry also results in large Rashba splitting on the order of tenths of eV.

\end{abstract}

\pacs{} 

\maketitle

Non-centrosymmetric superconductors have been known for decades, but have become a prominent research topic recently with the discovery of the heavy fermion superconductor CePt$_3$Si\cite{CePt3Si}. Non-centrosymmetric systems can exhibit asymmetric spin-orbit coupling (SOC) in superconducting materials, which leads to the breaking of spin degeneracy and a parity-mixed superconducting state\cite{bauer2012non}. The symmetry of the Cooper pairs is therefore nontrivially affected by the strength of the SOC, which is governed by the crystal structure as well as the elemental composition\cite{Matano},\cite{RashbaBiTeI}. Superconductors that lack inversion symmetry can be divided into two types: strongly correlated systems such as CePt$_3$Si\cite{CePt3Si} and UIr\cite{akazawa2004pressure}, and weakly-correlated systems such as Li$_2$M$_3$B (M = Pd, Pt)\cite{Togano1},\cite{Togano2} and Mg$_{10}$Ir$_{19}$B$_{16}$\cite{Mg10Ir19B16}. In the strongly correlated materials, the superconducting properties are heavily influenced by the electron correlations making the weakly correlated materials more fertile ground for studying the effects derived from the breaking of inversion symmetry and the asymmetric spin orbit coupling interaction. Materials with strong SOC are also of interest as exotic spin systems\cite{hyperkagome} and topological insulators\cite{TIHasan},\cite{zhang2009topological} (TIs), and relativistic Dirac electrons in condensed matter systems are of interest on the surface of TIs, in graphene and other monatomic hexagonal lattices\cite{novoselov2004electric},\cite{novoselov2005two},\cite{geim2007rise},\cite{yan2012prediction}, and also in the bulk of 3D Dirac semimetals (e.g. Cd$_3$As$_2$, Na$_3$Bi and Pb$_{1-x}$Sn$_x$Se\cite{borisenko2013experimental},\cite{neupane2013observation},\cite{liu2013discovery},\cite{liang2013evidence}). Recent theoretical work has predicted SOC gapping in Dirac cones at the K point in crystallographic phases similar to, but heavier, than graphene\cite{liu2011quantum}. Here we show that the hexagonal, non-centrosymmetric compound PbTaSe$_2$ exhibits strong SOC, superconducts below 3.72 K, and has a gapped graphene-like Dirac cone at K in its electronic structure that is derived exclusively from its hexagonal Pb layer.

\indent{}PbTaSe$_2$ displays alternating stacking of hexagonal TaSe$_2$ and Pb layers (Figure 1a)\cite{eppinga1980generalized}. High-quality polycrystalline samples were synthesized by solid state reaction at 800 $^\circ$C for one week using pre-reacted PbSe, TaSe$_2$, and elemental Ta powder in sealed quartz tubes in a PbSe atmosphere. The samples were determined to be pure by powder x-ray diffraction. Due to the polycrystalline nature of the samples, the measured superconductivity parameters are averaged over all crystallographic directions.   

\indent{}The superconducting transition was examined through temperature dependent measurements of the electrical resistivity ($\rho$(T)) and \textit{dc} magnetic susceptibility using a Quantum Design PPMS. The whole temperature range of $\rho$(T) is shown in Figure 1a. The normal state resistivity for PbTaSe$_2$ reveals metallic like character (d$\rho$/dT $~>~$ 0), with the residual resistivity ratio (RRR)~$\approx$~6. The superconducting transition is seen just below 3.8 K. Figure 1(b) shows the superconducting transition characterized by \textit{dc} magnetic susceptibility. The much smaller field-cooling (FC) signal compared to the zero-field-cooling (ZFC) signal, is caused by pinning of the vortices. The estimated superconducting critical temperature is 3.7 K, in agreement with resistivity measurement. The diamagnetic response, normalized by a demagnetization factor, is very close to the expected value. Figure 1c shows the low temperature resistivity $\rho$(T) under zero field and applied magnetic fields up to 0.5 T. A very sharp superconducting transition is observed for 0 T, with the superconducting critical temperature T$_c$ = 3.79 K and transition width $\Delta$T$_c$ = 0.15 K. Knowing the values of T$_c$ for different magnetic fields, we plot the upper critical field values, $\mu_0$H$_{c2}$ vs. temperature in Figure 1d. By using the Werthamer-Helfand-Hohenberg (WHH) relationship\cite{werthamer1966temperature}, we estimate the zero-temperature upper critical field $\mu_0$H$_{c2}$(0) = -0.7T$_c$ dH$_{c2}$/dT$_c$ = 1.17 T. With this information, the coherence length can be calculated by using the Ginzburg-Landau formula $\xi_{GL}$(0) = ($\phi_0$/2$\pi$H$_{c2}$(0))$^{1/2}$, where $\phi_0$=h/2e and is found to be $\xi_{GL}$(0) = 17 nm. 
\begin{figure}[t]
	\includegraphics[width=0.45\textwidth]{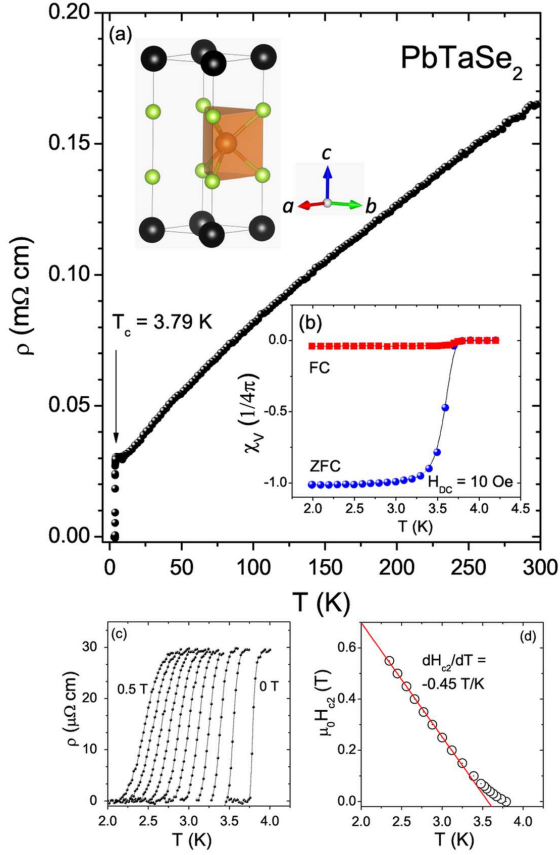}
	\caption{\scriptsize{\textbf{(color online):} (a) Resistivity as a function of temperature showing the superconducting transition for PbTaSe$_2$ at 3.79 K.  Inset: The crystal structure of PbTaSe$_2$, where Pb are the large black spheres, Se are the small green spheres, and Ta are the medium orange spheres. (b) shows the observed zero-field cooling (ZFC) and field cooling (FC) magnetic susceptibility measurements under magnetic field H$_{DC}$ = 10 Oe. The superconducting critical temperature (T$_c$) estimated from these measurements is 3.72 K. The maximum ZFC susceptibility is estimated to be -1.01(1/4$\pi$). (c) shows the superconducting transition under various magnetic fields and (d) plots the upper critical field values vs. temperature. The red solid line through the data shows the best linear fit with the initial slope dH$_{c2}$/dT = -0.45 T/K.}}
	\label{Figure_1}
\end{figure}

\indent{}Assuming that the initial linear response to field is perfectly diamagnetic, (Figure 2a) i.e. dM/dH is $-\frac{1}{4\pi}$, we obtain a demagnetization factor that is consistent with the sample’s shape and its orientation in the magnetic field. Figure 2a shows the magnetization (M) as a function of applied field (H) and Figure 2b presents the difference between magnetization measured at 2 K and the M$_{fit}$ (shown as a red solid line in Figure 2a) fitted in the low H range where the linear M(H) is observed. As shown in Figure 2b, M(H) starts to deviate from M$_{fit}$ at a field, H$_c^*$, of about 53 Oe, giving a lower critical field, taking into account the demagnetization factor, of H$_{c1}$(2K) = H$_c^*$/(1-d) = 60 Oe. The estimation of $\mu_0$H$_{c1}$(0) has been done by fitting experimental data to the formula H$_{c1}$(T) = H$_{c1}$(0)[1-(T/T$_c$)$^2$], which is represented by the red solid line in Figure 2c. The estimated zero-temperature lower critical field $\mu_0$H$_{c1}$(0) = 75 Oe, implies a Ginzburg-Landau superconducting penetration depth (calculated using $\mu_0H_{c1} = \frac{\Phi_0}{4\pi\lambda^2_{GL}}ln\frac{\lambda_{GL}}{\xi_{GL}}$ ) of approximately $\lambda_{GL}$ = 242 nm. The Ginzburg-Landau parameter ($\kappa = \lambda_{GL}(0) / \xi_{GL}(0)$) is $\kappa$ = 14, indicating that PbTaSe$_2$ is a type-II superconductor. Using these parameters and the relation H$_{c1}\cdot$H$_{c2}$ = H$_c^2$$\cdot$ln($\kappa$), we estimated the thermodynamic critical field H$_c$ = 574 Oe. 
\begin{figure}[b]
	\includegraphics[width=0.45\textwidth]{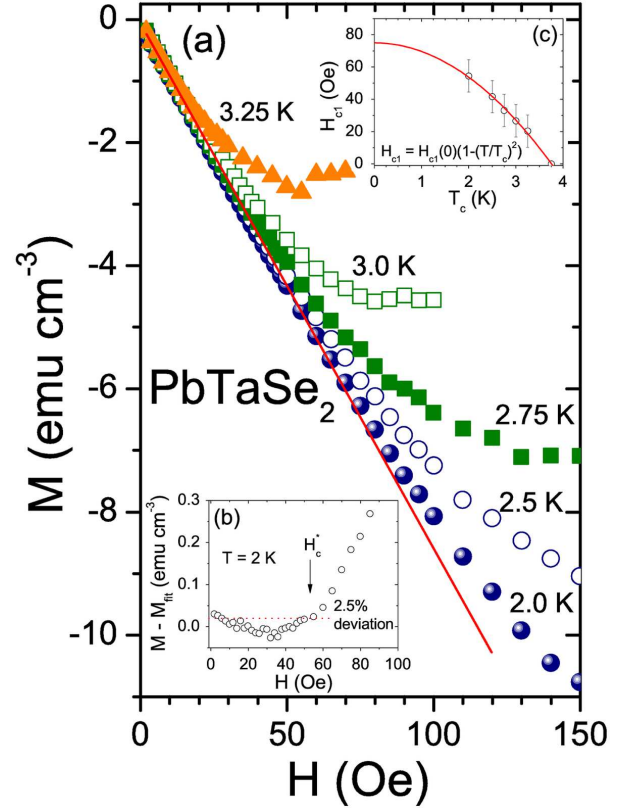}
	\caption{\scriptsize{\textbf{(color online):} (a) M vs. H for PbTaSe$_2$ at various temperatures. The solid red line is fitted to the 2 K data in the low H range where the linear M(H) is observed. (b) The difference between magnetization measured at 2 K and the M$_{fit}$:  the difference M(H) - M$_{fit}$ deviates more than 2.5\% above the fitted curve for H$_c^*$ $\approx$ 53 Oe. (c) The estimation of $\mu_0$H$_{c1}$(0) done by fitting the H$_{c1}$ data to the formula H$_{c1}$(T) = H$_{c1}$(0)[1-(T/T$_c$)$^2$], which is represented by the red solid line.}}
	\label{Figure_2}
\end{figure}

\indent{}The heat capacity was measured using a relaxation calorimeter (Quantum Design PPMS). Figure 3a illustrates the overall temperature dependence of the specific heat (C$_p$). At room temperature, C$_p$ is close to the expected Dulong-Petit value (3nR $\approx$ 100 J mol$^{-1 }$K$^{-1)}$, where n is the number of atoms per formula unit (n = 4), and R is the gas constant (R = 8.314 J mol$^{-1)}$ K$^{-1)}$). Figure 3b shows C$_p$/T versus T$^2$ in the low temperature range measured under a magnetic field of $\mu_0$H = 5 T, which exceeds the upper critical field for PbTaSe$_2$. The experimental data points were fitted in the temperature range of 1.9 K - 3.7 K, using the formula C$_p$ = $\gamma$T+$\beta$T$^3$. The fit yields the electronic specific heat coefficient (Sommerfeld coefficient) $\gamma$ = 6.9(2) mJ mol$^{-1}$ K$^{-2}$, and phonon specific heat coefficient $\beta$ = 2.67(0.03) mJ mol$^{-1}$ K$^{-4}$. Bulk superconductivity is confirmed by a large, anomaly (Figure 3c) at a temperature that is consistent with the T$_c$ determined by the \textit{dc} magnetic susceptibility and resistivity measurements. Using $\gamma$ and the specific heat jump value ($\Delta$C/T$_c$) at the superconducting transition temperature, $\Delta$C/$\gamma$T$_c$ can be calculated and is found to be 1.41, which is very close to the BCS value of 1.426. 
\begin{table}[t]
\caption{\footnotesize{Superconducting Parameters of PbTaSe$_2$}}
\scalebox{0.85}{
\begin{tabular}{lcccc}
\hline
\hline
Parameter & Unit & Nb$_{0.18}$Re$_{0.82}$\cite{Nb0.18Re0.82}& Mg$_{10}$Ir$_{19}$B$_{16}$\cite{Mg10Ir19B16}& PbTaSe$_2$ \\[1ex]
\hline 

T$_c$ & K & 8.8 & 4.45 & 3.72 \\ 
$\mu_0$H$_{c1}$(0) & Oe & 55.7 & 30 & 75  \\ 
$\mu_0$H$_{c2}$(0) & T & 17.3 & 0.77 & 1.17  \\ 
$\xi_{GL}$(0) & nm & 4.4 & 21 & 17  \\ 
$\lambda_{GL}$(0) & nm & 363 & 404 & 242  \\ 
$\kappa$(0) && 83 & 20 &  14 \\ 
$\gamma$(0) & mJ/molK$^2$ & 53.4 & 52.6 & 6.9  \\ 
$\Delta$C/$\gamma$T$_c$ && 1.86 & 1.60 & 1.41  \\ 
$\mu_0$H$^{Pauli}$ & T & 16.8 & 8.2 & 6.8  \\ 
$\Theta_D$ & K & 383 & 280 & 112  \\ 
$\lambda_{ep}$ && 0.73 & 0.66 & 0.74  \\ 

\hline
\hline
\end{tabular}}
\label{}
\end{table}
\indent{}In a simple Debye model, the $\beta$ coefficient is related to the Debye temperature ($\Theta_D$) through $\Theta_D = \left(\frac{12\pi^4}{5\beta}nR\right)^\frac{1}{3}$, and the estimated Debye temperature for PbTaSe$_2$ is only 143 K, which reflects the fact that it contains heavy elements. As can be seen from Figure 3a, C$_{Debye}$ with $\Theta_D$ = 143 K (solid blue line) is not large enough to reach the experimental heat capacity values above 40 K. Therefore we fitted the data in the temperature range 10 K - 300 K by using the following formula: C$_p$ = $\gamma$T + \textit{k}C$_{Debye}(T) + (1-\textit{k})C_{Einstein}(T)$ , in which higher energy optical modes are considered. The first term ($\gamma$T) is the electronic contribution and the \textit{k} parameter corresponds to the weight of the phonon contributions to the specific heat given by Debye (C$_{Debye}$) and Einstein (C$_{Einstein}$) models respectively:
 $C_{Debeye}(T) = 9nR\left(\frac{T}{\Theta_D}\right)^3\int\frac{x^4\exp (x)}{[\exp (x-1)]^2}$, and $C_{Einstein}(T) = 3nR\left(\frac{\Theta_E}{T}\right)^2\exp \left(\frac{\Theta_E}{T}\right)\left(\exp (\frac{\Theta_E}{T}-1)\right)^{-2}$. $\Theta_D$ and $\Theta_E$ are the Debye and Einstein temperatures respectively. The fit represented by solid, red line in Figure 3a, gives 41\% of the weight to a Debye term with $\Theta_D$ = 112 K, close to the temperature derived from the low temperature fit, and the remaining weight (59\%) in an Einstein mode with energy $\Theta_E$ = 290 K. 

\indent{}With these results and assuming $\mu^*$ = 0.13, the electron-phonon coupling constant ($\lambda_{ep}$) can be calculated from the inverted McMillan’s formula\cite{mcmillan1968transition}: $\lambda_{ep} = \frac{1.04~+~\mu^*\ln \left(\frac{\Theta_D}{1.45T_c}\right)}{\left(1-0.62\right)\mu^*\ln \left(\frac{\Theta_D}{1.45T_c}\right)~-~1.04}$ and is found to be 0.74. This value is similar to that found in other moderately coupled superconductors such as YPd$_2$Sn and HfPd$_2$Al\cite{klimczuk2012superconductivity}. Having the Sommerfeld parameter and the electron-phonon coupling, the non-interacting density of states at the Fermi energy can be calculated from: \textit{N}(E$_F$) = $\frac{3\gamma}{\pi^2k_B^2\left(1 + \lambda_{ep}\right)}$ The value obtained for PbTaSe$_2$, \textit{N}(E$_F$) = 1.7 states eV$^{-1}$ per formula unit, agrees well with the ~1.5 states eV$^{-1}$ per formula unit calculated from theoretical predictions (see below). Table 1 compares the measured and derived superconductivity parameters of PbTaSe$_2$ with other non-centrosymmetric superconductors, Mg$_{10}$Ir$_{19}$B$_{16}$ and Nb$_{0.18}$Re$_{0.82}$.
\begin{figure}[b]
	\includegraphics[width=0.45\textwidth]{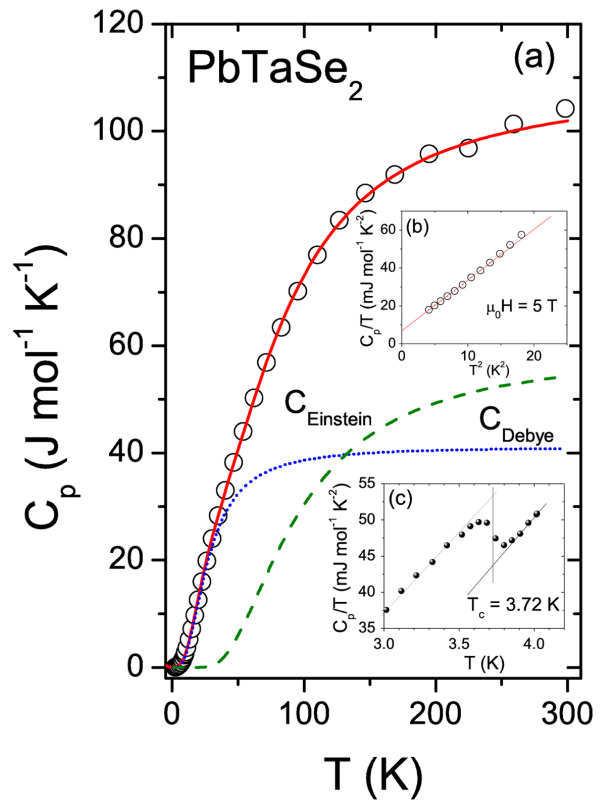}
	\caption{\scriptsize{\textbf{(color online):} (a) The specific heat versus temperature measurements represented by the open circles. The solid red line is a fit to a combined model where 41\% of the weight is given to the Debye model (dotted blue line) and 59\% to the Einstein model (dashed green line). (b) C$_p$/T versus T$^2$ with the red line showing the linear fit in the low temperature region. (c) C$_p$/T versus T showing the bulk superconducting jump and the equal area approximation for the T$_c$ determination. }}
	\label{Figure_3}
\end{figure}

\indent{}Electronic structure calculations were performed in the framework of density functional theory using the Wien2k code\cite{blaha1990full} with a full-potential linearized augmented plane-wave and local orbitals basis together with the Perdew-Burke-Ernzerhof parameterization of the generalized gradient approximation\cite{perdew1996generalized},\cite{CalcDetail}. In order to check the robustness of the electronic structure calculations, they were also performed using the Trans-Blaha modified Becke-Johnson (mBJ) functional\cite{tran2009accurate}, which resulted in no significant differences.

\indent{}Band structure calculations for PbTaSe$_2$ immediately unveil a single bulk 3D Dirac cone at the K point in the Brillouin zone that is gapped by large SOC (Figure 4a). The Dirac cone at K in graphene is also gapped by SOC, albeit by only a few mK; in PbTaSe$_2$, the strong SOC gaps the cone by about 0.8 eV. This single Dirac cone that is gapped by SOC is highly suggestive of PbTaSe$_2$ being topologically nontrivial, as this motif is observed in both graphene\cite{kane2005quantum} (which is a quantum spin hall insulator) and Bi$_{14}$Rh$_3$I$_9$\cite{rasche2013stacked} (which is predicted to be a weak topological insulator). Also, 3D Dirac cones have recently been observed in the two semi-metals Cd$_3$As$_2$ and Na$_3$Bi which, if gapped, would drive the systems into the topological insulator regime. Furthermore, the four states that were degenerate at the Dirac point without SOC, when SOC is included, now have different eigenvalues under S6 (C6-bar), C3, and mirror operations. The existence of these symmetries at points where there are possible band inversions suggests the possibility of topological surface states protected by crystalline symmetry, as is seen in Topological Crystalline Insulators\cite{dziawa2012topological},\cite{fu2011topological},\cite{hsieh2012topological}. Furthermore, although PbTaSe$_2$ is a metal, there is a continuous gap formed around E$_F$ when SOC is considered. A closer look at the electronic structure in Figure 4a reveals two band crossings along A-L and A-H (that are not present along Gamma-M or Gamma-K, demonstrating that the coupling in the z direction is important) that become gapped with the inclusion of SOC.

\indent{}The combination of large SOC and broken inversion symmetry is also apparent in the large spin splitting observed in the electronic structure. This is most readily observed around the H point in Figure 4a. This spin splitting is on a similar magnitude (in the tenths of eVs) as in the giant Rashba semiconductor BiTeI\cite{ishizaka2011giant}. In fact, Rashba type spin splitting is observed around the M and L points around the continuous gap shown in Figure 4a. 
\begin{figure}[h]
	\includegraphics[width=0.5\textwidth]{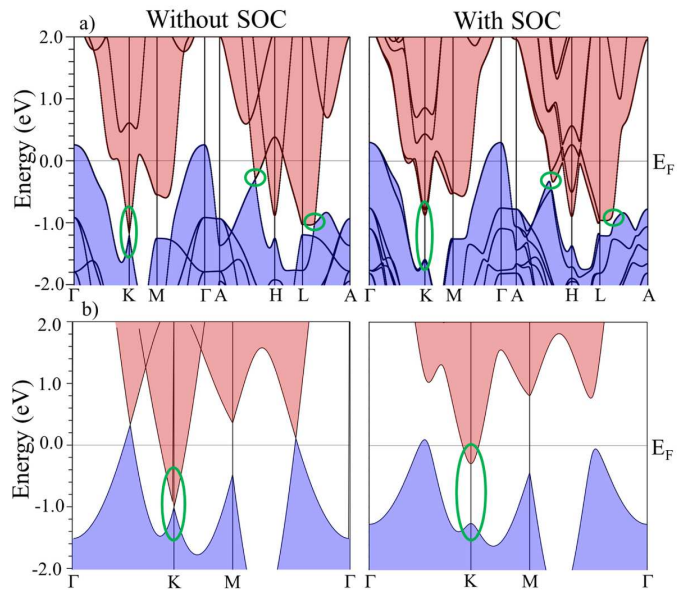}
	\caption{\scriptsize{\textbf{(color online):} (a) Calculated electronic structures of PbTaSe$_2$ with and without spin orbit coupling (SOC). Bands are shaded to highlight the continuous gap opened when SOC is included. The Dirac cone that is gapped with SOC as well as band crossings along A-H and A-L, which are similarly gapped implying a possible band inversion, are circled in green. (b) Electronic structure of the 2D Pb sublattice alone in PbTaSe$_2$, with green circles highlighting the Dirac cone that is gapped by SOC at K, just as in bulk PbTaSe$_2$.}}
	\label{Figure_4}
\end{figure}
Finally, Figure 4b shows the electronic structure of the Pb sublattice alone. This demonstrates that the Dirac cone observed at K is uniquely due to the Pb sublattice and, with SOC considered, becomes gapped as well. In fact, the Pb sublattice goes from being metallic to being almost completely gapped with the inclusion of SOC. The electronic structure therefore shows that some of the charge carriers in PbTaSe$_2$ are massive 3D Dirac electrons. If the apparent band inversion of the 3D massive Dirac electrons\cite{zhu2011field} gives rise to a topological crystalline insulator-like state as in Pb$_{0.77}$Sn$_{0.23}$Se\cite{dziawa2012topological}, then a cleaved 001 surface of PbTaSe$_2$ may host Majorana zero modes at the surface even if the bulk superconducting gap is nontrivial, due to the fact that the 001 surface maintains the mirror and C3 symmetries\cite{fang2013new}. Furthermore, the large spin splitting observed in the electronic structure indicates the likelihood of an unconventional pairing mechanism that could lead to a nontrivial superconducting gap - another possible way of supporting Majorana fermions at a cleaved surface of PbTaSe$_2$.

\indent{}In conclusion, we report the discovery of superconductivity in PbTaSe$_2$ and its unusual electronic structure. The single-layer Pb sublattice in PbTaSe$_2$ behaves similarly to graphene monolayers in that it also generates a Dirac point at K, generating 3D massive Dirac fermions by large SOC.  Unlike graphene superlattices, however, the inclusion of Pb layers in a natural superlattice with TaSe$_2$ does not affect the in-plane orbitals of Pb that make the Dirac cone at K. This represents a unique case where a 2D elemental sublattice capable of generating 2D massive Dirac fermions can be interfaced with a transiton metal dichalcogenide to create a superconducting superlattice, generating 3D massive Dirac fermions and broken inversion symmetry, all in a thermodynamically stable material. In addition to PbTaSe$_2$ (T$_c$ = 3.72 K) and InTaS$_2$ (T$_c$~$\approx$~1.0 K\cite{DiSalvo19731922}), layered materials of this type\cite{eppinga1980generalized},\cite{ali2013synthesis} may represent a new family of materials where the interplay of noncentrosymmetric superconductivity and large SOC can lead to nontrivial electronic topologies.

This research was supported by the US Department of Energy, grant DE FG02-98-ER45706.

\bibliography{Lit}

\begin{thebibliography}{39}%
\makeatletter
\providecommand \@ifxundefined [1]{%
 \@ifx{#1\undefined}
}%
\providecommand \@ifnum [1]{%
 \ifnum #1\expandafter \@firstoftwo
 \else \expandafter \@secondoftwo
 \fi
}%
\providecommand \@ifx [1]{%
 \ifx #1\expandafter \@firstoftwo
 \else \expandafter \@secondoftwo
 \fi
}%
\providecommand \natexlab [1]{#1}%
\providecommand \enquote  [1]{``#1''}%
\providecommand \bibnamefont  [1]{#1}%
\providecommand \bibfnamefont [1]{#1}%
\providecommand \citenamefont [1]{#1}%
\providecommand \href@noop [0]{\@secondoftwo}%
\providecommand \href [0]{\begingroup \@sanitize@url \@href}%
\providecommand \@href[1]{\@@startlink{#1}\@@href}%
\providecommand \@@href[1]{\endgroup#1\@@endlink}%
\providecommand \@sanitize@url [0]{\catcode `\\12\catcode `\$12\catcode
  `\&12\catcode `\#12\catcode `\^12\catcode `\_12\catcode `\%12\relax}%
\providecommand \@@startlink[1]{}%
\providecommand \@@endlink[0]{}%
\providecommand \url  [0]{\begingroup\@sanitize@url \@url }%
\providecommand \@url [1]{\endgroup\@href {#1}{\urlprefix }}%
\providecommand \urlprefix  [0]{URL }%
\providecommand \Eprint [0]{\href }%
\providecommand \doibase [0]{http://dx.doi.org/}%
\providecommand \selectlanguage [0]{\@gobble}%
\providecommand \bibinfo  [0]{\@secondoftwo}%
\providecommand \bibfield  [0]{\@secondoftwo}%
\providecommand \translation [1]{[#1]}%
\providecommand \BibitemOpen [0]{}%
\providecommand \bibitemStop [0]{}%
\providecommand \bibitemNoStop [0]{.\EOS\space}%
\providecommand \EOS [0]{\spacefactor3000\relax}%
\providecommand \BibitemShut  [1]{\csname bibitem#1\endcsname}%
\let\auto@bib@innerbib\@empty
\bibitem [{\citenamefont {Bauer}\ \emph {et~al.}(2004)\citenamefont {Bauer},
  \citenamefont {Hilscher}, \citenamefont {Michor}, \citenamefont {Paul},
  \citenamefont {Scheidt}, \citenamefont {Gribanov}, \citenamefont {Seropegin},
  \citenamefont {No\"el}, \citenamefont {Sigrist},\ and\ \citenamefont
  {Rogl}}]{CePt3Si}%
  \BibitemOpen
  \bibfield  {author} {\bibinfo {author} {\bibfnamefont {E.}~\bibnamefont
  {Bauer}}, \bibinfo {author} {\bibfnamefont {G.}~\bibnamefont {Hilscher}},
  \bibinfo {author} {\bibfnamefont {H.}~\bibnamefont {Michor}}, \bibinfo
  {author} {\bibfnamefont {C.}~\bibnamefont {Paul}}, \bibinfo {author}
  {\bibfnamefont {E.~W.}\ \bibnamefont {Scheidt}}, \bibinfo {author}
  {\bibfnamefont {A.}~\bibnamefont {Gribanov}}, \bibinfo {author}
  {\bibfnamefont {Y.}~\bibnamefont {Seropegin}}, \bibinfo {author}
  {\bibfnamefont {H.}~\bibnamefont {No\"el}}, \bibinfo {author} {\bibfnamefont
  {M.}~\bibnamefont {Sigrist}}, \ and\ \bibinfo {author} {\bibfnamefont
  {P.}~\bibnamefont {Rogl}},\ }\href {\doibase 10.1103/PhysRevLett.92.027003}
  {\bibfield  {journal} {\bibinfo  {journal} {Phys. Rev. Lett.}\ }\textbf
  {\bibinfo {volume} {92}},\ \bibinfo {pages} {027003} (\bibinfo {year}
  {2004})}\BibitemShut {NoStop}%
\bibitem [{\citenamefont {Bauer}\ and\ \citenamefont
  {Sigrist}(2012)}]{bauer2012non}%
  \BibitemOpen
  \bibfield  {author} {\bibinfo {author} {\bibfnamefont {E.}~\bibnamefont
  {Bauer}}\ and\ \bibinfo {author} {\bibfnamefont {M.}~\bibnamefont
  {Sigrist}},\ }\href@noop {} {\emph {\bibinfo {title} {Non-centrosymmetric
  Superconductors: Introduction and Overview}}},\ Vol.\ \bibinfo {volume}
  {847}\ (\bibinfo  {publisher} {Springer},\ \bibinfo {year}
  {2012})\BibitemShut {NoStop}%
\bibitem [{\citenamefont {Matano}\ \emph {et~al.}(2013)\citenamefont {Matano},
  \citenamefont {Maeda}, \citenamefont {Sawaoka}, \citenamefont {Muro},
  \citenamefont {Takabatake}, \citenamefont {Joshi}, \citenamefont
  {Ramakrishnan}, \citenamefont {Kawashima}, \citenamefont {Akimitsu},\ and\
  \citenamefont {Zheng}}]{Matano}%
  \BibitemOpen
  \bibfield  {author} {\bibinfo {author} {\bibfnamefont {K.}~\bibnamefont
  {Matano}}, \bibinfo {author} {\bibfnamefont {S.}~\bibnamefont {Maeda}},
  \bibinfo {author} {\bibfnamefont {H.}~\bibnamefont {Sawaoka}}, \bibinfo
  {author} {\bibfnamefont {Y.}~\bibnamefont {Muro}}, \bibinfo {author}
  {\bibfnamefont {T.}~\bibnamefont {Takabatake}}, \bibinfo {author}
  {\bibfnamefont {B.}~\bibnamefont {Joshi}}, \bibinfo {author} {\bibfnamefont
  {S.}~\bibnamefont {Ramakrishnan}}, \bibinfo {author} {\bibfnamefont
  {K.}~\bibnamefont {Kawashima}}, \bibinfo {author} {\bibfnamefont
  {J.}~\bibnamefont {Akimitsu}}, \ and\ \bibinfo {author} {\bibfnamefont
  {G.}~\bibnamefont {Zheng}},\ }\href {\doibase 10.7566/JPSJ.82.084711}
  {\bibfield  {journal} {\bibinfo  {journal} {J. Phys. Soc. Jpn.}\ }\textbf
  {\bibinfo {volume} {82}},\ \bibinfo {pages} {084711} (\bibinfo {year}
  {2013})}\BibitemShut {NoStop}%
\bibitem [{\citenamefont {Bahramy}\ \emph {et~al.}(2011)\citenamefont
  {Bahramy}, \citenamefont {Arita},\ and\ \citenamefont
  {Nagaosa}}]{RashbaBiTeI}%
  \BibitemOpen
  \bibfield  {author} {\bibinfo {author} {\bibfnamefont {M.~S.}\ \bibnamefont
  {Bahramy}}, \bibinfo {author} {\bibfnamefont {R.}~\bibnamefont {Arita}}, \
  and\ \bibinfo {author} {\bibfnamefont {N.}~\bibnamefont {Nagaosa}},\ }\href
  {\doibase 10.1103/PhysRevB.84.041202} {\bibfield  {journal} {\bibinfo
  {journal} {Phys. Rev. B}\ }\textbf {\bibinfo {volume} {84}},\ \bibinfo
  {pages} {041202} (\bibinfo {year} {2011})}\BibitemShut {NoStop}%
\bibitem [{\citenamefont {Akazawa}\ \emph {et~al.}(2004)\citenamefont
  {Akazawa}, \citenamefont {Hidaka}, \citenamefont {Fujiwara}, \citenamefont
  {Kobayashi}, \citenamefont {Yamamoto}, \citenamefont {Haga}, \citenamefont
  {Settai},\ and\ \citenamefont {{\=O}nuki}}]{akazawa2004pressure}%
  \BibitemOpen
  \bibfield  {author} {\bibinfo {author} {\bibfnamefont {T.}~\bibnamefont
  {Akazawa}}, \bibinfo {author} {\bibfnamefont {H.}~\bibnamefont {Hidaka}},
  \bibinfo {author} {\bibfnamefont {T.}~\bibnamefont {Fujiwara}}, \bibinfo
  {author} {\bibfnamefont {T.}~\bibnamefont {Kobayashi}}, \bibinfo {author}
  {\bibfnamefont {E.}~\bibnamefont {Yamamoto}}, \bibinfo {author}
  {\bibfnamefont {Y.}~\bibnamefont {Haga}}, \bibinfo {author} {\bibfnamefont
  {R.}~\bibnamefont {Settai}}, \ and\ \bibinfo {author} {\bibfnamefont
  {Y.}~\bibnamefont {{\=O}nuki}},\ }\href@noop {} {\bibfield  {journal}
  {\bibinfo  {journal} {Journal of Physics: Condensed Matter}\ }\textbf
  {\bibinfo {volume} {16}},\ \bibinfo {pages} {L29} (\bibinfo {year}
  {2004})}\BibitemShut {NoStop}%
\bibitem [{\citenamefont {Togano}\ \emph {et~al.}(2004)\citenamefont {Togano},
  \citenamefont {Badica}, \citenamefont {Nakamori}, \citenamefont {Orimo},
  \citenamefont {Takeya},\ and\ \citenamefont {Hirata}}]{Togano1}%
  \BibitemOpen
  \bibfield  {author} {\bibinfo {author} {\bibfnamefont {K.}~\bibnamefont
  {Togano}}, \bibinfo {author} {\bibfnamefont {P.}~\bibnamefont {Badica}},
  \bibinfo {author} {\bibfnamefont {Y.}~\bibnamefont {Nakamori}}, \bibinfo
  {author} {\bibfnamefont {S.}~\bibnamefont {Orimo}}, \bibinfo {author}
  {\bibfnamefont {H.}~\bibnamefont {Takeya}}, \ and\ \bibinfo {author}
  {\bibfnamefont {K.}~\bibnamefont {Hirata}},\ }\href {\doibase
  10.1103/PhysRevLett.93.247004} {\bibfield  {journal} {\bibinfo  {journal}
  {Phys. Rev. Lett.}\ }\textbf {\bibinfo {volume} {93}},\ \bibinfo {pages}
  {247004} (\bibinfo {year} {2004})}\BibitemShut {NoStop}%
\bibitem [{\citenamefont {Badica}\ \emph {et~al.}(2005)\citenamefont {Badica},
  \citenamefont {Kondo},\ and\ \citenamefont {Togano}}]{Togano2}%
  \BibitemOpen
  \bibfield  {author} {\bibinfo {author} {\bibfnamefont {P.}~\bibnamefont
  {Badica}}, \bibinfo {author} {\bibfnamefont {T.}~\bibnamefont {Kondo}}, \
  and\ \bibinfo {author} {\bibfnamefont {K.}~\bibnamefont {Togano}},\ }\href
  {\doibase 10.1143/JPSJ.74.1014} {\bibfield  {journal} {\bibinfo  {journal}
  {Journal of the Physical Society of Japan}\ }\textbf {\bibinfo {volume}
  {74}},\ \bibinfo {pages} {1014} (\bibinfo {year} {2005})}\BibitemShut
  {NoStop}%
\bibitem [{\citenamefont {Klimczuk}\ \emph {et~al.}(2007)\citenamefont
  {Klimczuk}, \citenamefont {Ronning}, \citenamefont {Sidorov}, \citenamefont
  {Cava},\ and\ \citenamefont {Thompson}}]{Mg10Ir19B16}%
  \BibitemOpen
  \bibfield  {author} {\bibinfo {author} {\bibfnamefont {T.}~\bibnamefont
  {Klimczuk}}, \bibinfo {author} {\bibfnamefont {F.}~\bibnamefont {Ronning}},
  \bibinfo {author} {\bibfnamefont {V.}~\bibnamefont {Sidorov}}, \bibinfo
  {author} {\bibfnamefont {R.~J.}\ \bibnamefont {Cava}}, \ and\ \bibinfo
  {author} {\bibfnamefont {J.~D.}\ \bibnamefont {Thompson}},\ }\href {\doibase
  10.1103/PhysRevLett.99.257004} {\bibfield  {journal} {\bibinfo  {journal}
  {Phys. Rev. Lett.}\ }\textbf {\bibinfo {volume} {99}},\ \bibinfo {pages}
  {257004} (\bibinfo {year} {2007})}\BibitemShut {NoStop}%
\bibitem [{\citenamefont {Okamoto}\ \emph {et~al.}(2007)\citenamefont
  {Okamoto}, \citenamefont {Nohara}, \citenamefont {Aruga-Katori},\ and\
  \citenamefont {Takagi}}]{hyperkagome}%
  \BibitemOpen
  \bibfield  {author} {\bibinfo {author} {\bibfnamefont {Y.}~\bibnamefont
  {Okamoto}}, \bibinfo {author} {\bibfnamefont {M.}~\bibnamefont {Nohara}},
  \bibinfo {author} {\bibfnamefont {H.}~\bibnamefont {Aruga-Katori}}, \ and\
  \bibinfo {author} {\bibfnamefont {H.}~\bibnamefont {Takagi}},\ }\href
  {\doibase 10.1103/PhysRevLett.99.137207} {\bibfield  {journal} {\bibinfo
  {journal} {Phys. Rev. Lett.}\ }\textbf {\bibinfo {volume} {99}},\ \bibinfo
  {pages} {137207} (\bibinfo {year} {2007})}\BibitemShut {NoStop}%
\bibitem [{\citenamefont {Hasan}\ and\ \citenamefont {Kane}(2010)}]{TIHasan}%
  \BibitemOpen
  \bibfield  {author} {\bibinfo {author} {\bibfnamefont {M.~Z.}\ \bibnamefont
  {Hasan}}\ and\ \bibinfo {author} {\bibfnamefont {C.~L.}\ \bibnamefont
  {Kane}},\ }\href {\doibase 10.1103/RevModPhys.82.3045} {\bibfield  {journal}
  {\bibinfo  {journal} {Rev. Mod. Phys.}\ }\textbf {\bibinfo {volume} {82}},\
  \bibinfo {pages} {3045} (\bibinfo {year} {2010})}\BibitemShut {NoStop}%
\bibitem [{\citenamefont {Zhang}\ \emph {et~al.}(2009)\citenamefont {Zhang},
  \citenamefont {Liu}, \citenamefont {Qi}, \citenamefont {Dai}, \citenamefont
  {Fang},\ and\ \citenamefont {Zhang}}]{zhang2009topological}%
  \BibitemOpen
  \bibfield  {author} {\bibinfo {author} {\bibfnamefont {H.}~\bibnamefont
  {Zhang}}, \bibinfo {author} {\bibfnamefont {C.-X.}\ \bibnamefont {Liu}},
  \bibinfo {author} {\bibfnamefont {X.-L.}\ \bibnamefont {Qi}}, \bibinfo
  {author} {\bibfnamefont {X.}~\bibnamefont {Dai}}, \bibinfo {author}
  {\bibfnamefont {Z.}~\bibnamefont {Fang}}, \ and\ \bibinfo {author}
  {\bibfnamefont {S.-C.}\ \bibnamefont {Zhang}},\ }\href@noop {} {\bibfield
  {journal} {\bibinfo  {journal} {Nature Physics}\ }\textbf {\bibinfo {volume}
  {5}},\ \bibinfo {pages} {438} (\bibinfo {year} {2009})}\BibitemShut {NoStop}%
\bibitem [{\citenamefont {Novoselov}\ \emph {et~al.}(2004)\citenamefont
  {Novoselov}, \citenamefont {Geim}, \citenamefont {Morozov}, \citenamefont
  {Jiang}, \citenamefont {Zhang}, \citenamefont {Dubonos}, \citenamefont
  {Grigorieva},\ and\ \citenamefont {Firsov}}]{novoselov2004electric}%
  \BibitemOpen
  \bibfield  {author} {\bibinfo {author} {\bibfnamefont {K.~S.}\ \bibnamefont
  {Novoselov}}, \bibinfo {author} {\bibfnamefont {A.~K.}\ \bibnamefont {Geim}},
  \bibinfo {author} {\bibfnamefont {S.}~\bibnamefont {Morozov}}, \bibinfo
  {author} {\bibfnamefont {D.}~\bibnamefont {Jiang}}, \bibinfo {author}
  {\bibfnamefont {Y.}~\bibnamefont {Zhang}}, \bibinfo {author} {\bibfnamefont
  {S.}~\bibnamefont {Dubonos}}, \bibinfo {author} {\bibfnamefont
  {I.}~\bibnamefont {Grigorieva}}, \ and\ \bibinfo {author} {\bibfnamefont
  {A.}~\bibnamefont {Firsov}},\ }\href@noop {} {\bibfield  {journal} {\bibinfo
  {journal} {Science}\ }\textbf {\bibinfo {volume} {306}},\ \bibinfo {pages}
  {666} (\bibinfo {year} {2004})}\BibitemShut {NoStop}%
\bibitem [{\citenamefont {Novoselov}\ \emph {et~al.}(2005)\citenamefont
  {Novoselov}, \citenamefont {Geim}, \citenamefont {Morozov}, \citenamefont
  {Jiang}, \citenamefont {Grigorieva}, \citenamefont {Dubonos},\ and\
  \citenamefont {Firsov}}]{novoselov2005two}%
  \BibitemOpen
  \bibfield  {author} {\bibinfo {author} {\bibfnamefont {K.}~\bibnamefont
  {Novoselov}}, \bibinfo {author} {\bibfnamefont {A.~K.}\ \bibnamefont {Geim}},
  \bibinfo {author} {\bibfnamefont {S.}~\bibnamefont {Morozov}}, \bibinfo
  {author} {\bibfnamefont {D.}~\bibnamefont {Jiang}}, \bibinfo {author}
  {\bibfnamefont {M.~K.~I.}\ \bibnamefont {Grigorieva}}, \bibinfo {author}
  {\bibfnamefont {S.}~\bibnamefont {Dubonos}}, \ and\ \bibinfo {author}
  {\bibfnamefont {A.}~\bibnamefont {Firsov}},\ }\href@noop {} {\bibfield
  {journal} {\bibinfo  {journal} {nature}\ }\textbf {\bibinfo {volume} {438}},\
  \bibinfo {pages} {197} (\bibinfo {year} {2005})}\BibitemShut {NoStop}%
\bibitem [{\citenamefont {Geim}\ and\ \citenamefont
  {Novoselov}(2007)}]{geim2007rise}%
  \BibitemOpen
  \bibfield  {author} {\bibinfo {author} {\bibfnamefont {A.~K.}\ \bibnamefont
  {Geim}}\ and\ \bibinfo {author} {\bibfnamefont {K.~S.}\ \bibnamefont
  {Novoselov}},\ }\href@noop {} {\bibfield  {journal} {\bibinfo  {journal}
  {Nature materials}\ }\textbf {\bibinfo {volume} {6}},\ \bibinfo {pages} {183}
  (\bibinfo {year} {2007})}\BibitemShut {NoStop}%
\bibitem [{\citenamefont {Yan}\ \emph {et~al.}(2012)\citenamefont {Yan},
  \citenamefont {M{\"u}chler},\ and\ \citenamefont
  {Felser}}]{yan2012prediction}%
  \BibitemOpen
  \bibfield  {author} {\bibinfo {author} {\bibfnamefont {B.}~\bibnamefont
  {Yan}}, \bibinfo {author} {\bibfnamefont {L.}~\bibnamefont {M{\"u}chler}}, \
  and\ \bibinfo {author} {\bibfnamefont {C.}~\bibnamefont {Felser}},\
  }\href@noop {} {\bibfield  {journal} {\bibinfo  {journal} {Physical Review
  Letters}\ }\textbf {\bibinfo {volume} {109}},\ \bibinfo {pages} {116406}
  (\bibinfo {year} {2012})}\BibitemShut {NoStop}%
\bibitem [{\citenamefont {Borisenko}\ \emph {et~al.}(2013)\citenamefont
  {Borisenko}, \citenamefont {Gibson}, \citenamefont {Evtushinsky},
  \citenamefont {Zabolotnyy}, \citenamefont {Buechner},\ and\ \citenamefont
  {Cava}}]{borisenko2013experimental}%
  \BibitemOpen
  \bibfield  {author} {\bibinfo {author} {\bibfnamefont {S.}~\bibnamefont
  {Borisenko}}, \bibinfo {author} {\bibfnamefont {Q.}~\bibnamefont {Gibson}},
  \bibinfo {author} {\bibfnamefont {D.}~\bibnamefont {Evtushinsky}}, \bibinfo
  {author} {\bibfnamefont {V.}~\bibnamefont {Zabolotnyy}}, \bibinfo {author}
  {\bibfnamefont {B.}~\bibnamefont {Buechner}}, \ and\ \bibinfo {author}
  {\bibfnamefont {R.~J.}\ \bibnamefont {Cava}},\ }\href@noop {} {\bibfield
  {journal} {\bibinfo  {journal} {arXiv preprint arXiv:1309.7978}\ } (\bibinfo
  {year} {2013})}\BibitemShut {NoStop}%
\bibitem [{\citenamefont {Neupane}\ \emph {et~al.}(2013)\citenamefont
  {Neupane}, \citenamefont {Xu}, \citenamefont {Sankar}, \citenamefont
  {Alidoust}, \citenamefont {Bian}, \citenamefont {Liu}, \citenamefont
  {Belopolski}, \citenamefont {Chang}, \citenamefont {Jeng}, \citenamefont
  {Lin} \emph {et~al.}}]{neupane2013observation}%
  \BibitemOpen
  \bibfield  {author} {\bibinfo {author} {\bibfnamefont {M.}~\bibnamefont
  {Neupane}}, \bibinfo {author} {\bibfnamefont {S.}~\bibnamefont {Xu}},
  \bibinfo {author} {\bibfnamefont {R.}~\bibnamefont {Sankar}}, \bibinfo
  {author} {\bibfnamefont {N.}~\bibnamefont {Alidoust}}, \bibinfo {author}
  {\bibfnamefont {G.}~\bibnamefont {Bian}}, \bibinfo {author} {\bibfnamefont
  {C.}~\bibnamefont {Liu}}, \bibinfo {author} {\bibfnamefont {I.}~\bibnamefont
  {Belopolski}}, \bibinfo {author} {\bibfnamefont {T.-R.}\ \bibnamefont
  {Chang}}, \bibinfo {author} {\bibfnamefont {H.-T.}\ \bibnamefont {Jeng}},
  \bibinfo {author} {\bibfnamefont {H.}~\bibnamefont {Lin}},  \emph {et~al.},\
  }\href@noop {} {\bibfield  {journal} {\bibinfo  {journal} {arXiv preprint
  arXiv:1309.7892}\ } (\bibinfo {year} {2013})}\BibitemShut {NoStop}%
\bibitem [{\citenamefont {Liu}\ \emph {et~al.}(2013)\citenamefont {Liu},
  \citenamefont {Zhou}, \citenamefont {Wang}, \citenamefont {Weng},
  \citenamefont {Prabhakaran}, \citenamefont {Mo}, \citenamefont {Zhang},
  \citenamefont {Shen}, \citenamefont {Fang}, \citenamefont {Dai} \emph
  {et~al.}}]{liu2013discovery}%
  \BibitemOpen
  \bibfield  {author} {\bibinfo {author} {\bibfnamefont {Z.}~\bibnamefont
  {Liu}}, \bibinfo {author} {\bibfnamefont {B.}~\bibnamefont {Zhou}}, \bibinfo
  {author} {\bibfnamefont {Z.}~\bibnamefont {Wang}}, \bibinfo {author}
  {\bibfnamefont {H.}~\bibnamefont {Weng}}, \bibinfo {author} {\bibfnamefont
  {D.}~\bibnamefont {Prabhakaran}}, \bibinfo {author} {\bibfnamefont {S.-K.}\
  \bibnamefont {Mo}}, \bibinfo {author} {\bibfnamefont {Y.}~\bibnamefont
  {Zhang}}, \bibinfo {author} {\bibfnamefont {Z.}~\bibnamefont {Shen}},
  \bibinfo {author} {\bibfnamefont {Z.}~\bibnamefont {Fang}}, \bibinfo {author}
  {\bibfnamefont {X.}~\bibnamefont {Dai}},  \emph {et~al.},\ }\href@noop {}
  {\bibfield  {journal} {\bibinfo  {journal} {arXiv preprint arXiv:1310.0391}\
  } (\bibinfo {year} {2013})}\BibitemShut {NoStop}%
\bibitem [{\citenamefont {Liang}\ \emph {et~al.}(2013)\citenamefont {Liang},
  \citenamefont {Gibson}, \citenamefont {Xiong}, \citenamefont {Hirschberger},
  \citenamefont {Koduvayur}, \citenamefont {Cava},\ and\ \citenamefont
  {Ong}}]{liang2013evidence}%
  \BibitemOpen
  \bibfield  {author} {\bibinfo {author} {\bibfnamefont {T.}~\bibnamefont
  {Liang}}, \bibinfo {author} {\bibfnamefont {Q.}~\bibnamefont {Gibson}},
  \bibinfo {author} {\bibfnamefont {J.}~\bibnamefont {Xiong}}, \bibinfo
  {author} {\bibfnamefont {M.}~\bibnamefont {Hirschberger}}, \bibinfo {author}
  {\bibfnamefont {S.~P.}\ \bibnamefont {Koduvayur}}, \bibinfo {author}
  {\bibfnamefont {R.}~\bibnamefont {Cava}}, \ and\ \bibinfo {author}
  {\bibfnamefont {N.}~\bibnamefont {Ong}},\ }\href@noop {} {\bibfield
  {journal} {\bibinfo  {journal} {arXiv preprint arXiv:1307.4022}\ } (\bibinfo
  {year} {2013})}\BibitemShut {NoStop}%
\bibitem [{\citenamefont {Liu}\ \emph {et~al.}(2011)\citenamefont {Liu},
  \citenamefont {Feng},\ and\ \citenamefont {Yao}}]{liu2011quantum}%
  \BibitemOpen
  \bibfield  {author} {\bibinfo {author} {\bibfnamefont {C.-C.}\ \bibnamefont
  {Liu}}, \bibinfo {author} {\bibfnamefont {W.}~\bibnamefont {Feng}}, \ and\
  \bibinfo {author} {\bibfnamefont {Y.}~\bibnamefont {Yao}},\ }\href@noop {}
  {\bibfield  {journal} {\bibinfo  {journal} {Physical review letters}\
  }\textbf {\bibinfo {volume} {107}},\ \bibinfo {pages} {076802} (\bibinfo
  {year} {2011})}\BibitemShut {NoStop}%
\bibitem [{\citenamefont {Eppinga}\ and\ \citenamefont
  {Wiegers}(1980)}]{eppinga1980generalized}%
  \BibitemOpen
  \bibfield  {author} {\bibinfo {author} {\bibfnamefont {R.}~\bibnamefont
  {Eppinga}}\ and\ \bibinfo {author} {\bibfnamefont {G.}~\bibnamefont
  {Wiegers}},\ }\href@noop {} {\bibfield  {journal} {\bibinfo  {journal}
  {Physica B+ C}\ }\textbf {\bibinfo {volume} {99}},\ \bibinfo {pages} {121}
  (\bibinfo {year} {1980})}\BibitemShut {NoStop}%
\bibitem [{\citenamefont {Werthamer}\ \emph {et~al.}(1966)\citenamefont
  {Werthamer}, \citenamefont {Helfand},\ and\ \citenamefont
  {Hohenberg}}]{werthamer1966temperature}%
  \BibitemOpen
  \bibfield  {author} {\bibinfo {author} {\bibfnamefont {N.}~\bibnamefont
  {Werthamer}}, \bibinfo {author} {\bibfnamefont {E.}~\bibnamefont {Helfand}},
  \ and\ \bibinfo {author} {\bibfnamefont {P.}~\bibnamefont {Hohenberg}},\
  }\href@noop {} {\bibfield  {journal} {\bibinfo  {journal} {Physical Review}\
  }\textbf {\bibinfo {volume} {147}},\ \bibinfo {pages} {295} (\bibinfo {year}
  {1966})}\BibitemShut {NoStop}%
\bibitem [{\citenamefont {Karki}\ \emph {et~al.}(2011)\citenamefont {Karki},
  \citenamefont {Xiong}, \citenamefont {Haldolaarachchige}, \citenamefont
  {Stadler}, \citenamefont {Vekhter}, \citenamefont {Adams}, \citenamefont
  {Young}, \citenamefont {Phelan},\ and\ \citenamefont {Chan}}]{Nb0.18Re0.82}%
  \BibitemOpen
  \bibfield  {author} {\bibinfo {author} {\bibfnamefont {A.~B.}\ \bibnamefont
  {Karki}}, \bibinfo {author} {\bibfnamefont {Y.~M.}\ \bibnamefont {Xiong}},
  \bibinfo {author} {\bibfnamefont {N.}~\bibnamefont {Haldolaarachchige}},
  \bibinfo {author} {\bibfnamefont {S.}~\bibnamefont {Stadler}}, \bibinfo
  {author} {\bibfnamefont {I.}~\bibnamefont {Vekhter}}, \bibinfo {author}
  {\bibfnamefont {P.~W.}\ \bibnamefont {Adams}}, \bibinfo {author}
  {\bibfnamefont {D.~P.}\ \bibnamefont {Young}}, \bibinfo {author}
  {\bibfnamefont {W.~A.}\ \bibnamefont {Phelan}}, \ and\ \bibinfo {author}
  {\bibfnamefont {J.~Y.}\ \bibnamefont {Chan}},\ }\href {\doibase
  10.1103/PhysRevB.83.144525} {\bibfield  {journal} {\bibinfo  {journal} {Phys.
  Rev. B}\ }\textbf {\bibinfo {volume} {83}},\ \bibinfo {pages} {144525}
  (\bibinfo {year} {2011})}\BibitemShut {NoStop}%
\bibitem [{\citenamefont {McMillan}(1968)}]{mcmillan1968transition}%
  \BibitemOpen
  \bibfield  {author} {\bibinfo {author} {\bibfnamefont {W.}~\bibnamefont
  {McMillan}},\ }\href@noop {} {\bibfield  {journal} {\bibinfo  {journal}
  {Physical Review}\ }\textbf {\bibinfo {volume} {167}},\ \bibinfo {pages}
  {331} (\bibinfo {year} {1968})}\BibitemShut {NoStop}%
\bibitem [{\citenamefont {Klimczuk}\ \emph {et~al.}(2012)\citenamefont
  {Klimczuk}, \citenamefont {Wang}, \citenamefont {Gofryk}, \citenamefont
  {Ronning}, \citenamefont {Winterlik}, \citenamefont {Fecher}, \citenamefont
  {Griveau}, \citenamefont {Colineau}, \citenamefont {Felser}, \citenamefont
  {Thompson} \emph {et~al.}}]{klimczuk2012superconductivity}%
  \BibitemOpen
  \bibfield  {author} {\bibinfo {author} {\bibfnamefont {T.}~\bibnamefont
  {Klimczuk}}, \bibinfo {author} {\bibfnamefont {C.}~\bibnamefont {Wang}},
  \bibinfo {author} {\bibfnamefont {K.}~\bibnamefont {Gofryk}}, \bibinfo
  {author} {\bibfnamefont {F.}~\bibnamefont {Ronning}}, \bibinfo {author}
  {\bibfnamefont {J.}~\bibnamefont {Winterlik}}, \bibinfo {author}
  {\bibfnamefont {G.}~\bibnamefont {Fecher}}, \bibinfo {author} {\bibfnamefont
  {J.-C.}\ \bibnamefont {Griveau}}, \bibinfo {author} {\bibfnamefont
  {E.}~\bibnamefont {Colineau}}, \bibinfo {author} {\bibfnamefont
  {C.}~\bibnamefont {Felser}}, \bibinfo {author} {\bibfnamefont
  {J.}~\bibnamefont {Thompson}},  \emph {et~al.},\ }\href@noop {} {\bibfield
  {journal} {\bibinfo  {journal} {Physical Review B}\ }\textbf {\bibinfo
  {volume} {85}},\ \bibinfo {pages} {174505} (\bibinfo {year}
  {2012})}\BibitemShut {NoStop}%
\bibitem [{\citenamefont {Blaha}\ \emph {et~al.}(1990)\citenamefont {Blaha},
  \citenamefont {Schwarz}, \citenamefont {Sorantin},\ and\ \citenamefont
  {Trickey}}]{blaha1990full}%
  \BibitemOpen
  \bibfield  {author} {\bibinfo {author} {\bibfnamefont {P.}~\bibnamefont
  {Blaha}}, \bibinfo {author} {\bibfnamefont {K.}~\bibnamefont {Schwarz}},
  \bibinfo {author} {\bibfnamefont {P.}~\bibnamefont {Sorantin}}, \ and\
  \bibinfo {author} {\bibfnamefont {S.}~\bibnamefont {Trickey}},\ }\href@noop
  {} {\bibfield  {journal} {\bibinfo  {journal} {Computer Physics
  Communications}\ }\textbf {\bibinfo {volume} {59}},\ \bibinfo {pages} {399}
  (\bibinfo {year} {1990})}\BibitemShut {NoStop}%
\bibitem [{\citenamefont {Perdew}\ \emph {et~al.}(1996)\citenamefont {Perdew},
  \citenamefont {Burke},\ and\ \citenamefont
  {Ernzerhof}}]{perdew1996generalized}%
  \BibitemOpen
  \bibfield  {author} {\bibinfo {author} {\bibfnamefont {J.~P.}\ \bibnamefont
  {Perdew}}, \bibinfo {author} {\bibfnamefont {K.}~\bibnamefont {Burke}}, \
  and\ \bibinfo {author} {\bibfnamefont {M.}~\bibnamefont {Ernzerhof}},\
  }\href@noop {} {\bibfield  {journal} {\bibinfo  {journal} {Physical review
  letters}\ }\textbf {\bibinfo {volume} {77}},\ \bibinfo {pages} {3865}
  (\bibinfo {year} {1996})}\BibitemShut {NoStop}%
\bibitem [{Cal()}]{CalcDetail}%
  \BibitemOpen
  \href@noop {} {\ }\bibinfo {note} {The plane wave cutoff parameter RMTKmax
  was set to 8 and the Brilloun zone (BZ) was sampled by 10000k-points, or in
  the case of the Pb sublattice calculation, a 3x3x1 k-mesh}\BibitemShut
  {NoStop}%
\bibitem [{\citenamefont {Tran}\ and\ \citenamefont
  {Blaha}(2009)}]{tran2009accurate}%
  \BibitemOpen
  \bibfield  {author} {\bibinfo {author} {\bibfnamefont {F.}~\bibnamefont
  {Tran}}\ and\ \bibinfo {author} {\bibfnamefont {P.}~\bibnamefont {Blaha}},\
  }\href@noop {} {\bibfield  {journal} {\bibinfo  {journal} {Physical review
  letters}\ }\textbf {\bibinfo {volume} {102}},\ \bibinfo {pages} {226401}
  (\bibinfo {year} {2009})}\BibitemShut {NoStop}%
\bibitem [{\citenamefont {Kane}\ and\ \citenamefont
  {Mele}(2005)}]{kane2005quantum}%
  \BibitemOpen
  \bibfield  {author} {\bibinfo {author} {\bibfnamefont {C.}~\bibnamefont
  {Kane}}\ and\ \bibinfo {author} {\bibfnamefont {E.}~\bibnamefont {Mele}},\
  }\href@noop {} {\bibfield  {journal} {\bibinfo  {journal} {Physical Review
  Letters}\ }\textbf {\bibinfo {volume} {95}},\ \bibinfo {pages} {226801}
  (\bibinfo {year} {2005})}\BibitemShut {NoStop}%
\bibitem [{\citenamefont {Rasche}\ \emph {et~al.}(2013)\citenamefont {Rasche},
  \citenamefont {Isaeva}, \citenamefont {Ruck}, \citenamefont {Borisenko},
  \citenamefont {Zabolotnyy}, \citenamefont {B{\"u}chner}, \citenamefont
  {Koepernik}, \citenamefont {Ortix}, \citenamefont {Richter},\ and\
  \citenamefont {van~den Brink}}]{rasche2013stacked}%
  \BibitemOpen
  \bibfield  {author} {\bibinfo {author} {\bibfnamefont {B.}~\bibnamefont
  {Rasche}}, \bibinfo {author} {\bibfnamefont {A.}~\bibnamefont {Isaeva}},
  \bibinfo {author} {\bibfnamefont {M.}~\bibnamefont {Ruck}}, \bibinfo {author}
  {\bibfnamefont {S.}~\bibnamefont {Borisenko}}, \bibinfo {author}
  {\bibfnamefont {V.}~\bibnamefont {Zabolotnyy}}, \bibinfo {author}
  {\bibfnamefont {B.}~\bibnamefont {B{\"u}chner}}, \bibinfo {author}
  {\bibfnamefont {K.}~\bibnamefont {Koepernik}}, \bibinfo {author}
  {\bibfnamefont {C.}~\bibnamefont {Ortix}}, \bibinfo {author} {\bibfnamefont
  {M.}~\bibnamefont {Richter}}, \ and\ \bibinfo {author} {\bibfnamefont
  {J.}~\bibnamefont {van~den Brink}},\ }\href@noop {} {\bibfield  {journal}
  {\bibinfo  {journal} {Nature materials}\ } (\bibinfo {year}
  {2013})}\BibitemShut {NoStop}%
\bibitem [{\citenamefont {Dziawa}\ \emph {et~al.}(2012)\citenamefont {Dziawa},
  \citenamefont {Kowalski}, \citenamefont {Dybko}, \citenamefont {Buczko},
  \citenamefont {Szczerbakow}, \citenamefont {Szot}, \citenamefont
  {{\L}usakowska}, \citenamefont {Balasubramanian}, \citenamefont {Wojek},
  \citenamefont {Berntsen} \emph {et~al.}}]{dziawa2012topological}%
  \BibitemOpen
  \bibfield  {author} {\bibinfo {author} {\bibfnamefont {P.}~\bibnamefont
  {Dziawa}}, \bibinfo {author} {\bibfnamefont {B.}~\bibnamefont {Kowalski}},
  \bibinfo {author} {\bibfnamefont {K.}~\bibnamefont {Dybko}}, \bibinfo
  {author} {\bibfnamefont {R.}~\bibnamefont {Buczko}}, \bibinfo {author}
  {\bibfnamefont {A.}~\bibnamefont {Szczerbakow}}, \bibinfo {author}
  {\bibfnamefont {M.}~\bibnamefont {Szot}}, \bibinfo {author} {\bibfnamefont
  {E.}~\bibnamefont {{\L}usakowska}}, \bibinfo {author} {\bibfnamefont
  {T.}~\bibnamefont {Balasubramanian}}, \bibinfo {author} {\bibfnamefont
  {B.~M.}\ \bibnamefont {Wojek}}, \bibinfo {author} {\bibfnamefont
  {M.}~\bibnamefont {Berntsen}},  \emph {et~al.},\ }\href@noop {} {\bibfield
  {journal} {\bibinfo  {journal} {Nature Materials}\ }\textbf {\bibinfo
  {volume} {11}},\ \bibinfo {pages} {1023} (\bibinfo {year}
  {2012})}\BibitemShut {NoStop}%
\bibitem [{\citenamefont {Fu}(2011)}]{fu2011topological}%
  \BibitemOpen
  \bibfield  {author} {\bibinfo {author} {\bibfnamefont {L.}~\bibnamefont
  {Fu}},\ }\href@noop {} {\bibfield  {journal} {\bibinfo  {journal} {Physical
  Review Letters}\ }\textbf {\bibinfo {volume} {106}},\ \bibinfo {pages}
  {106802} (\bibinfo {year} {2011})}\BibitemShut {NoStop}%
\bibitem [{\citenamefont {Hsieh}\ \emph {et~al.}(2012)\citenamefont {Hsieh},
  \citenamefont {Lin}, \citenamefont {Liu}, \citenamefont {Duan}, \citenamefont
  {Bansil},\ and\ \citenamefont {Fu}}]{hsieh2012topological}%
  \BibitemOpen
  \bibfield  {author} {\bibinfo {author} {\bibfnamefont {T.~H.}\ \bibnamefont
  {Hsieh}}, \bibinfo {author} {\bibfnamefont {H.}~\bibnamefont {Lin}}, \bibinfo
  {author} {\bibfnamefont {J.}~\bibnamefont {Liu}}, \bibinfo {author}
  {\bibfnamefont {W.}~\bibnamefont {Duan}}, \bibinfo {author} {\bibfnamefont
  {A.}~\bibnamefont {Bansil}}, \ and\ \bibinfo {author} {\bibfnamefont
  {L.}~\bibnamefont {Fu}},\ }\href@noop {} {\bibfield  {journal} {\bibinfo
  {journal} {Nature Communications}\ }\textbf {\bibinfo {volume} {3}},\
  \bibinfo {pages} {982} (\bibinfo {year} {2012})}\BibitemShut {NoStop}%
\bibitem [{\citenamefont {Ishizaka}\ \emph {et~al.}(2011)\citenamefont
  {Ishizaka}, \citenamefont {Bahramy}, \citenamefont {Murakawa}, \citenamefont
  {Sakano}, \citenamefont {Shimojima}, \citenamefont {Sonobe}, \citenamefont
  {Koizumi}, \citenamefont {Shin}, \citenamefont {Miyahara}, \citenamefont
  {Kimura} \emph {et~al.}}]{ishizaka2011giant}%
  \BibitemOpen
  \bibfield  {author} {\bibinfo {author} {\bibfnamefont {K.}~\bibnamefont
  {Ishizaka}}, \bibinfo {author} {\bibfnamefont {M.}~\bibnamefont {Bahramy}},
  \bibinfo {author} {\bibfnamefont {H.}~\bibnamefont {Murakawa}}, \bibinfo
  {author} {\bibfnamefont {M.}~\bibnamefont {Sakano}}, \bibinfo {author}
  {\bibfnamefont {T.}~\bibnamefont {Shimojima}}, \bibinfo {author}
  {\bibfnamefont {T.}~\bibnamefont {Sonobe}}, \bibinfo {author} {\bibfnamefont
  {K.}~\bibnamefont {Koizumi}}, \bibinfo {author} {\bibfnamefont
  {S.}~\bibnamefont {Shin}}, \bibinfo {author} {\bibfnamefont {H.}~\bibnamefont
  {Miyahara}}, \bibinfo {author} {\bibfnamefont {A.}~\bibnamefont {Kimura}},
  \emph {et~al.},\ }\href@noop {} {\bibfield  {journal} {\bibinfo  {journal}
  {Nature materials}\ }\textbf {\bibinfo {volume} {10}},\ \bibinfo {pages}
  {521} (\bibinfo {year} {2011})}\BibitemShut {NoStop}%
\bibitem [{\citenamefont {Zhu}\ \emph {et~al.}(2011)\citenamefont {Zhu},
  \citenamefont {Collaudin}, \citenamefont {Fauqu{\'e}}, \citenamefont {Kang},\
  and\ \citenamefont {Behnia}}]{zhu2011field}%
  \BibitemOpen
  \bibfield  {author} {\bibinfo {author} {\bibfnamefont {Z.}~\bibnamefont
  {Zhu}}, \bibinfo {author} {\bibfnamefont {A.}~\bibnamefont {Collaudin}},
  \bibinfo {author} {\bibfnamefont {B.}~\bibnamefont {Fauqu{\'e}}}, \bibinfo
  {author} {\bibfnamefont {W.}~\bibnamefont {Kang}}, \ and\ \bibinfo {author}
  {\bibfnamefont {K.}~\bibnamefont {Behnia}},\ }\href@noop {} {\bibfield
  {journal} {\bibinfo  {journal} {Nature Physics}\ }\textbf {\bibinfo {volume}
  {8}},\ \bibinfo {pages} {89} (\bibinfo {year} {2011})}\BibitemShut {NoStop}%
\bibitem [{\citenamefont {Fang}\ \emph {et~al.}(2013)\citenamefont {Fang},
  \citenamefont {Gilbert},\ and\ \citenamefont {Bernevig}}]{fang2013new}%
  \BibitemOpen
  \bibfield  {author} {\bibinfo {author} {\bibfnamefont {C.}~\bibnamefont
  {Fang}}, \bibinfo {author} {\bibfnamefont {M.~J.}\ \bibnamefont {Gilbert}}, \
  and\ \bibinfo {author} {\bibfnamefont {B.~A.}\ \bibnamefont {Bernevig}},\
  }\href@noop {} {\bibfield  {journal} {\bibinfo  {journal} {arXiv preprint
  arXiv:1308.2424}\ } (\bibinfo {year} {2013})}\BibitemShut {NoStop}%
\bibitem [{\citenamefont {Di~Salvo}\ \emph {et~al.}(1973)\citenamefont
  {Di~Salvo}, \citenamefont {Hull~Jr.}, \citenamefont {Schwartz}, \citenamefont
  {Voorhoeve},\ and\ \citenamefont {Waszczak}}]{DiSalvo19731922}%
  \BibitemOpen
  \bibfield  {author} {\bibinfo {author} {\bibfnamefont {F.}~\bibnamefont
  {Di~Salvo}}, \bibinfo {author} {\bibfnamefont {G.}~\bibnamefont {Hull~Jr.}},
  \bibinfo {author} {\bibfnamefont {L.}~\bibnamefont {Schwartz}}, \bibinfo
  {author} {\bibfnamefont {J.}~\bibnamefont {Voorhoeve}}, \ and\ \bibinfo
  {author} {\bibfnamefont {J.}~\bibnamefont {Waszczak}},\ }\href
  {http://www.scopus.com/inward/record.url?eid=2-s2.0-0000317785&partnerID=40&md5=3a31dceb7d64200ab85f9635d6682e27}
  {\bibfield  {journal} {\bibinfo  {journal} {The Journal of Chemical Physics}\
  }\textbf {\bibinfo {volume} {59}},\ \bibinfo {pages} {1922} (\bibinfo {year}
  {1973})}\BibitemShut {NoStop}%
\bibitem [{\citenamefont {Ali}\ \emph {et~al.}(2013)\citenamefont {Ali},
  \citenamefont {Ji}, \citenamefont {Hirai}, \citenamefont {Fuccillo},\ and\
  \citenamefont {Cava}}]{ali2013synthesis}%
  \BibitemOpen
  \bibfield  {author} {\bibinfo {author} {\bibfnamefont {M.~N.}\ \bibnamefont
  {Ali}}, \bibinfo {author} {\bibfnamefont {H.}~\bibnamefont {Ji}}, \bibinfo
  {author} {\bibfnamefont {D.}~\bibnamefont {Hirai}}, \bibinfo {author}
  {\bibfnamefont {M.}~\bibnamefont {Fuccillo}}, \ and\ \bibinfo {author}
  {\bibfnamefont {R.}~\bibnamefont {Cava}},\ }\href@noop {} {\bibfield
  {journal} {\bibinfo  {journal} {Journal of Solid State Chemistry}\ }
  (\bibinfo {year} {2013})}\BibitemShut {NoStop}%
\end{thebibliography}%

\end{document}